\documentclass[12pt,preprint]{aastex}

\shorttitle{Star Formation in RCW49}
\shortauthors{B. Whitney et al.}

\begin{document}

\title{A GLIMPSE of Star Formation in the Giant H II Region RCW 49}


\author{B. A. Whitney,\altaffilmark{1} R. Indebetouw,\altaffilmark{2}
B. L. Babler,\altaffilmark{2}
M. R. Meade,\altaffilmark{2} C. Watson,\altaffilmark{2}
M. J. Wolff,\altaffilmark{1} M. G. Wolfire,\altaffilmark{3}
D. P. Clemens,\altaffilmark{4} 
T. M. Bania,\altaffilmark{4} R. A. Benjamin,\altaffilmark{5}
M. Cohen,\altaffilmark{6}
K. E. Devine,\altaffilmark{2} J. M. Dickey,\altaffilmark{7}
F. Heitsch,\altaffilmark{8}
J. M. Jackson,\altaffilmark{4} 
H. A. Kobulnicky,\altaffilmark{9} A. P. Marston,\altaffilmark{10}
J. S. Mathis,\altaffilmark{2} E. P. Mercer,\altaffilmark{4}
J. R. Stauffer,\altaffilmark{11} S. R. Stolovy,\altaffilmark{11}
E. Churchwell\altaffilmark{2}
}





\altaffiltext{1}{Space Science Institute, 4750 Walnut St. Suite 205,
Boulder, CO 80301}

\altaffiltext{2}{University of Wisconsin-Madison, Dept. of Astronomy,
475 N. Charter St., Madison, WI 53706} 

\altaffiltext{3}{University of Maryland, Dept. Astronomy, College
Park, MD 20742-2421}

\altaffiltext{4}{Boston University, Institute for Astrophysical
Research, 725 Commonwealth Ave., Boston, MA 02215}

\altaffiltext{5}{University of Wisconsin-Whitewater, Physics Dept.,
800 W. Main St., Whitewater, WI 53190}

\altaffiltext{6}{University of California-Berkeley, Radio Astronomy
Lab, 601 Campbell Hall, Berkeley, CA 94720}

\altaffiltext{7}{University of Minnesota, Dept. Astronomy, 116 Church
St., SE, Minneapolis, MN 55455}

\altaffiltext{8}{Institute for Astronomy \& Astrophysics, University of Munich,
Scheinerstrasse 1, 81679 Munich
}

\altaffiltext{9}{University of Wyoming, Dept. Physics \& Astronomy, PO
Box 3905, Laramie, WY 82072}

\altaffiltext{10}{ESTEC/SCI-SA,Postbus 299,2200 AG Noordwijk,The
Netherlands}

\altaffiltext{11}{Caltech, Spitzer Science Center, MS 314-6, Pasadena,
CA 91125}



\begin{abstract}

GLIMPSE imaging using the Infrared Array Camera (IRAC) on the
{\it Spitzer Space Telescope} indicates that star formation is ongoing
in the RCW~49 giant \ion{H}{2} region.
A photometric comparison of the sources in RCW~49 to a similar area to
its north finds that at least 300 stars brighter than 13th
magnitude in band [3.6] have infrared excesses inconsistent with reddening due to
foreground extinction.  
These are likely young stellar objects (YSOs)
more massive than 2.5 $M_{\sun}$ suggesting 
that thousands more low-mass stars
are forming in this cloud.  
Some of the YSOs are massive (B
stars) and therefore very young, suggesting that a new generation of
star formation is occurring, possibly triggered by stellar winds and
shocks generated by the older (2-3 Myr) central massive cluster.  
The {\it Spitzer}
IRAC camera
has proven to be ideally suited for distinguishing young stars from
field stars, and the GLIMPSE survey of the Galactic Plane
will likely find thousands of
new star formation regions. 

\end{abstract}





\section{Introduction}

RCW~49 is one of the most luminous and massive \ion{H}{2} regions in
the Galaxy.  At its center lies the Westerlund~2 (hereafter W2; Westerlund 1960)
compact cluster which
contains over a dozen OB stars and a Wolf-Rayet star; another Wolf-Rayet star
lies several arc minutes away in the extended nebula (Moffat \& Vogt
1975; Moffat, Shara, \& Potter 1991; Carraro \& Munari 2004; van der
Hucht 2001).  The age of the W2 cluster is estimated to be 2-3 Myr
(Piatti et al. 1998).  As discussed in Churchwell et al. (2004; hereafter, C04), distance estimates range from $\sim 2.5 - 8$ kpc and we follow
their adoption of 4.2 kpc.  At this distance, we estimate the
cluster luminosity is 1.4$\times10^7$~$L_{\sun}$ based
on the IRAS flux and a relationship derived between far-infrared and
bolometric luminosity \citep{hunter2000}.
The stellar mass is estimated to be 
$\sim3 \times 10^4~$M$_{\sun}$ from the
radio-derived ionizing flux \citep{goss70,vacca96}

The RCW~49 region was observed with the {\it Spitzer} (Werner et al. 2004) Infrared Array
Camera (IRAC; Fazio et al. 2004) as part of the $\underline{G}$alactic $\underline{L}$egacy
$\underline{I}$nfrared $\underline{M}$id-$\underline{P}$lane $\underline{S}$urvey 
$\underline{E}$xtraordinaire 
\citep[GLIMPSE][]{pasp}
observing strategy validation.  C04
presents IRAC images of this region that show highly-structured, extended
polycyclic aromatic hydrocarbon (PAH) and gas emission, 
extending out to $\sim12$\arcmin\ from the W2 cluster center.  
The PAH emission is likely excited by
the strong ultraviolet/optical radiation from the central W2 cluster
(Leger \& Puget 1984; Allamandola et al. 1989), suggesting that large regions of the cloud are optically
thin to the central cluster radiation.
Furthermore, radio and infrared imaging show that at least the
southwest part of the cloud is blown out \citep{whiteoak97}
(all directions in this paper are referred to in Galactic coordinates).
Without detailed studies of the molecular gas in the RCW 49 region, it is unknown
if dense cores of gas and dust remain or if the dust is optically thin.
Has star formation ceased and is the cloud in the process of disruption?
Or is star formation ongoing, perhaps triggered by winds and shocks from
the central cluster?
This paper presents IRAC photometry of the RCW~49 region which reveals several
hundred sources with large infrared excesses, likely indicating youth.
%

After discussing the observations in \S2, we present images of selected regions in \S3 and
color-color and color-magnitude diagrams in \S4.
In \S5 we show spectral energy distributions (SEDs) of two massive (B2) YSOs.
\S6 concludes with a discussion of the current generation of star
formation and how it may relate to the W2 cluster.

\section{Observations}

The observations are described in detail in C04.  
A 1.7$^{\circ}$x0.7$^{\circ}$ region was imaged ten times with
1.2~s exposures in the four IRAC bands.  We will refer to these bands
by their central wavelength in $\mu$m, i.e., [3.6], [4.5], [5.8], and [8.0].
The data were taken on 2003 December 23 (Spitzer Program ID 195), 
and were processed by the
{\it Spitzer} Science Center (SSC) Pipeline (version S9.0.1).  
Positional accuracies are better
than 1\arcsec\ (Werner et al. 2004).  Point source full-width-half-max resolutions range
from $\simeq$1.6\arcsec\ at [3.6] to $\simeq$1.9\arcsec\ at [8.0].
The data were further processed by the GLIMPSE pipeline \citep{pasp}:
Point sources were extracted from each frame using a modified version
of DAOPHOT (Stetson 1987), and cross-referenced using the SSC bandmerger.
%
We produced a catalog of all sources detected at least 8 out of 10 times in
any of the four IRAC bands.  
For this present study, we culled the catalog to include only those sources with
signal-to-noise greater than 10 in any band.
We estimate the photometric errors from simulations by
placing point sources on a realistic background
(using the RCW~49 diffuse flux after stars are subtracted
out) and comparing the processed photometry to the input
values. 
The root-mean-square errors are  $<$0.05 mag in bands [3.6] and [4.5] for sources
brighter than 13 and 12.5 magnitude, respectively; 
$<$0.07 mag in band [5.8] and $<$0.15 mag in band [8.0]
for sources brighter than 11.7 magnitude.
Only sources brighter than these magnitudes (in a given band)
are displayed in the color-color and color-magnitude diagrams in \S4.
The flux calibration was checked
against five early A-type dwarf stars, 
and agrees to within 7\% in all IRAC bands
with the calculated fluxes (Cohen et al. 2003, Kurucz 1993).  

\section{Zones of Star Formation}

Figure~\ref{3col} shows a 3-color image mosaic\footnote{Mosaics were created
using Montage software, funded by the National Aeronautics and Space Administration's Earth Science Technology Office.}
at K$_s$, [3.6], and [4.5] $\mu$m (the K$_s$-band image is a mosaic of 2MASS\footnote{The Two Micron All Sky Survey is a joint 
project of the University of Massachusetts and the Infrared Processing
and Analysis Center/California Institute of Technology, funded by the
National Aeronautics and Space Administration and the National Science
Foundation.} images).
The [3.6] band diffuse emission (in green) is dominated by PAHs and small grains; [4.5]
emission (red) is mostly hydrogen Br$\alpha$ and small grains (C04); and
K$_s$-band (blue) is likely Br$\gamma$ and perhaps dust scattering.
Main sequence stars appear blue in these images.
The boxed regions in Figure~\ref{3col} are shown to larger scale in Figure~\ref{regions}.

The top left panel of Figure~\ref{regions} shows the W2 cluster.  This cluster
contains five O7V stars, one O6V, and a WN7 star (Moffat et al. 1991).
The bright star about 1\arcmin\ northwest of the cluster marked by the arrow is an O7
supergiant (Moffat et al. 1991).  Winds from the W2 cluster have produced
for a radio ring about 4\arcmin\ in radius centered on the cluster \citep{whiteoak97}.
The second region (region 2) is 3.4\arcmin\ SW of the W2 center.  
As discussed in \S5, the bright red source and diffuse green source (marked)
are likely massive (B3) YSOs.  
This cluster resides (in projection) in the ``blow-out''
region of the RCW~49 nebula where the radio ring is disrupted and
winds and UV radiation are probably escaping (Whiteoak \& Uchida 1997;
C04, Figure~1).
The third highlighted region, at 4.3\arcmin\ SE of W2, contains a
column of material similar to the elephant trunks of M16
(Hester et al. 1996).  The red star at the tip of 
the elephant trunk is a strong IR
emitter; and the bright red star to the SE is likely a massive
YSO since it has similar flux and colors to the two sources
discussed in \S5.  To the north at the edge of the image lies the
second known Wolf-Rayet star in the RCW~49 nebula (Moffat et al. 1991). 
It is thought to be responsible for a second radio bubble \citep{whiteoak97}. 
The fourth region is located 7.7\arcmin\ SE of W2.  It
contains several faint, red sources and dark lanes.
Region~5 is located 7.8\arcmin\ E of W2, and
Region~6 is located 9.1\arcmin\ south of
W2 and each contains a small group of red sources.  Region~6 
 is located on the
perimeter of the blow-out region.

These images of clustered red sources and dark lanes
suggest that star formation is taking place
throughout the RCW~49 nebula.  
We next examine stellar photometry of the entire region to determine if this is indeed the case.

\section{Color-Color and Color-Magnitude Diagrams}

Figure~\ref{ccd} shows color-color plots of the RCW~49 region (green
circles), and a ``field'' region of similar area about 0.75\arcdeg\ to the north
(pink circles) centered at (l,b)=(284.18,0.41).  
The black cross at upper right in each panel shows typical error bars for
each color sequence.

The black arrows show reddening vectors derived from optical/near-IR
extinction studies of the diffuse ISM (Cardelli, Clayton, \& Mathis 1989), 
and the grey arrows show reddening vectors derived
from this dataset (Indebetouw et al. 2004, in preparation).  Note that
the length of the arrows in the top left panel represent an optical
extinction of A$_V = 3$ whereas the length of the arrows in the other
three panels show A$_V = 30$.  
The black symbols represent colors from models of
1 million year old sources 
(using Siess et al. 2000 evolutionary tracks for central
source temperatures and luminosity)
surrounded by varying amounts of circumstellar gas and dust ($10^{-4}$ - 10 M$_\sun$).
The models were computed using a radiative transfer code that calculates
emission from  envelopes, disks and outflows surrounding stars \citep{whitney03}.  
The model central
star masses are 2.5, 3.8, and 5.9~M$_\sun$, corresponding to luminosities
of 10~L$_\sun$ (filled circles), 100~L$_\sun$ (x's), and 730~L$_\sun$
(open triangles).   

The 2MASS catalog was used to create the near-IR color sequence in the top left
panel of Figure~\ref{ccd}.   Only stars brighter than 14th magnitude in all three bands
are plotted since the errors increase below this.
The near-IR color sequence is typical of a star formation
region. 
The field stars (pink) follow the reddening line.  The stars
that are very red or below the reddening line are usually classified as
IR excess sources with disks or envelopes.  However, many IR excess
sources lie along the reddening line and are difficult
to distinguish from reddened main sequence stars.  In contrast,
the other panels in Figure~\ref{ccd} using IRAC data show that infrared-excess sources
in RCW~49 are well separated from field stars, which generally have
a small color range.  

Comparing the colors of the RCW~49 stars to the field stars, we find
that 357 stars out of 1393 in RCW~49 have colors [3.6]-[4.5]$>0.2$ mag, [4.5]-[5.8]$>0.35$ mag, 
or [5.8]-[8.0]$>0.2$ mag compared to 21/1114 in the field (about 2\%).
We note that due to our conservative brightness limits for this initial study,
many of the faint stars apparent in the
images in Figure~\ref{regions} are not included in the
color-color plots.  
Even so, it is likely that at least 300 of the sources in
this dataset are surrounded by warm circumstellar dust which indicates that
they have recently formed or are in the process of formation.

Figure~\ref{cmd} shows color-magnitude diagrams of the RCW~49 region
(green circles) and the northern field (pink).  The model symbols
(black) are as described in Figure~\ref{ccd}.  
The width of the field star distribution matches reasonably well to the error bars at
upper right.
The horizontal dashed line in each panel shows the expected brightness
of a B5 main sequence star at a distance of 4.2 kpc (synthesized
from the 2-35$\mu$m spectra of Cohen, 1993).  
Based on the number of
YSOs more luminous than this (about 300), and assuming a
fairly standard IMF \citep[][p. 488]{allen}, 
we estimate the total number of YSOs to be $\sim$7000
and their total mass to be $\sim$4500~M$_{\sun}$.

Where is most of the star formation occuring in this cloud?  
Figure~\ref{radial} plots the fraction of red sources as a function of
radius from the W2 cluster center, where red is defined as
[3.6]-[4.5]$>0.2$ mag, [4.5]-[5.8]$>0.35$ mag, and [5.8]-[8.0]$>0.3$ mag
(based on the color-magnitude diagrams of Figure~\ref{cmd}).
This distribution peaks within about 4\arcmin\ ($\sim2$ pc) from the W2 cluster center,
and decreases beyond that.  
The diffuse emission has a similar distribution (C04, Figure 3) suggesting 
the possibility of a systematic bias towards finding red stars due to higher [5.8] and  [8.0]
backgrounds.  However, the
density of total stars (of all colors) as a function of radius is fairly constant beyond about
2\arcmin\, suggesting that we are seeing the entire population of
stars above our magnitude limits.  
The correlation between the red stars and the diffuse flux is probably real and
due to 
higher diffuse flux in regions
of higher star formation activity.
We conclude from this plot that star formation is occurring preferentially
within about 4\arcmin\ ($\sim5$ pc) from the W2 cluster center and decreases
beyond that. 
Interestingly, this is the approximate outer radius of the radio ring \citep{whiteoak97}
and the ``plateau region'' of the diffuse flux (C04, Figure 3) where it is thought
that winds and radiation have swepted up and compressed gas and dust.  This
may have produced conditions for a second generation of star formation.
Since the medium is not smooth, the induced star formation likely occurs
over a range of radii.  The YSOs in Region 2 (Figures 1 and 2) and the elephant
trunk in Region 3 fall within this region.  Interestingly, the YSOs in Region 6
lie on the edge of the blow-out cavity.  
However, Regions 4 and 5 are well beyond the radio ring and may represent a
more quiescent form of star formation.

\section{SEDs of selected Sources}

In this section, we present SEDs of 
just two of the many interesting sources in the RCW~49 nebula.  These
are in Region~2 of Figure~\ref{regions}, marked by arrows.
The observed SEDs include data from 2MASS, {\it Spitzer} IRAC, and
{\it MSX}\footnote{This publication makes use of data products from the
{\it Midcourse Space Experiment}.  Processing of the data was funded by the Ballistic Missile Defense Organization with additional support from NASA Office of Space Science}.  
The source indicated by the left arrow in Region~2 is the brightest source in the region,
and is mildly saturated at IRAC [5.8] and [8.0] bands.
The source shown by the right arrow is a triangular-shaped greenish object.  
The SEDs of both sources are shown in Figure~\ref{sed}.  
IRAC photometry of the bright source was done by our automated pipeline.  We did
aperture photometry of the triangular source over a 4\arcsec\ aperture.
The flux at $\lambda > 8 \mu$m from these sources is blended at {\it MSX} resolution (18\arcsec),
though likely dominated by the source at left.  

It is clear from the brightness of these sources that they are luminous and likely
high-mass.
We computed model SEDs for the YSOs
using the code of Whitney et al. (2003).  
Alvarez, Hoara, \& Lucas (2004) modeled near-IR scattered light images 
of massive YSOs using 
the same geometries we use to fit the SEDs in Figure~\ref{sed}:
rotationally flattened infalling envelopes, disks, and dust-filled outflow cavities.
The model that fits the bright source SED best is powered by a B2 main sequence
star, has an envelope
with an infall rate of $2.7 \times 10^{-5}$ M$_\sun$/yr and a disk radius
of 500 AU; this gives an envelope extinction along the line-of-sight of A$_V=30$.
The model that fits the triangular source is powered by a B2.5 star,
has an infall rate of $8 \times 10^{-5}$ M$_\sun$/yr, disk radius
of 100 AU, and envelope extinction of 200 along the line-of-sight.
Both models are viewed at an inclination of  $i=70$\arcdeg (where
pole-on is defined as $i=0$\arcdeg),
and have foreground extinctions of A$_V=$15.
The SEDs could probably be fit with other envelope structures as long as a substantial 
amount of dust resides close to the star, and therefore the
stars are sufficiently
young that very little dust clearing has occured.

\section{Conclusions}

\begin{figure}
\epsscale{1}
\plotone{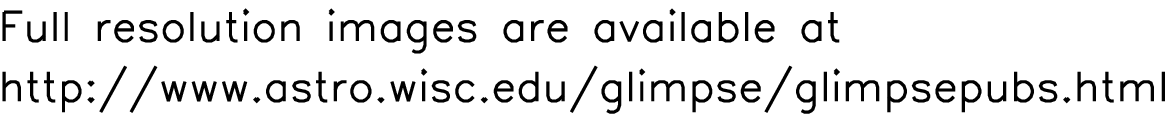}
\caption{\label{3col}
Three-color composite image of the RCW~49 region. The K$_s$ image is displayed as blue,
[3.6] as green, and [4.5] as red.  The images are scaled logarithmically.  
The blue boxes mark the regions shown in Figure~\ref{regions}.  
}
\end{figure}

\begin{figure}
\epsscale{1}
\plotone{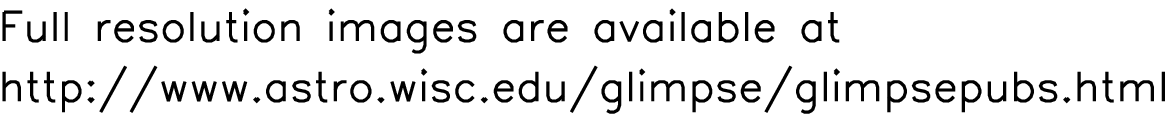}
\caption{\label{regions}
K$_s$-[3.6]-[4.5] 3-color composite images of selected regions (the marked boxes in Figure~\ref{3col}).
Each image is 3\arcmin\ on a side.  Arrows mark sources discussed in the text.}
\end{figure}

\begin{figure}
\epsscale{1}
\plotone{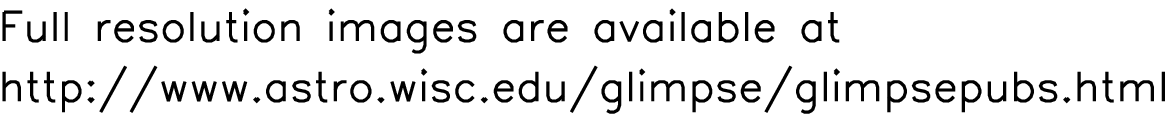}
\caption{\label{ccd} Color-color plots.  The black and grey arrows show the reddening
vectors for two different extinction laws (see text).  Sources from
the RCW~49 nebula are shown in green, and those from an off-cloud region
are shown in pink.
Typical error bars are shown at upper right of each panel.
YSO models of various
masses are also shown as black symbols (open triangles:  5.9 M$_\sun$; x's: 3.8 M$_\sun$; filled circles: 2.5 M$_\sun$).
}
\end{figure}

\begin{figure}
\plotone{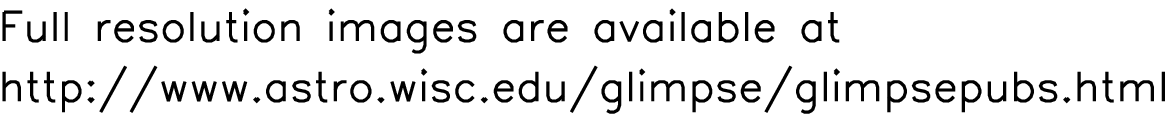} 
\caption{\label{cmd} Color-magnitude diagrams.  They symbols are as described in
Figure~\ref{ccd}. 
The dashed line shows the magnitude
of a B5 main sequence star at 4.2 kpc. 
The solid lines show the envelope created by the magnitude limits of the source list.
Typical error bars are shown at top right of each panel.
}
\end{figure}

\begin{figure}
\plotone{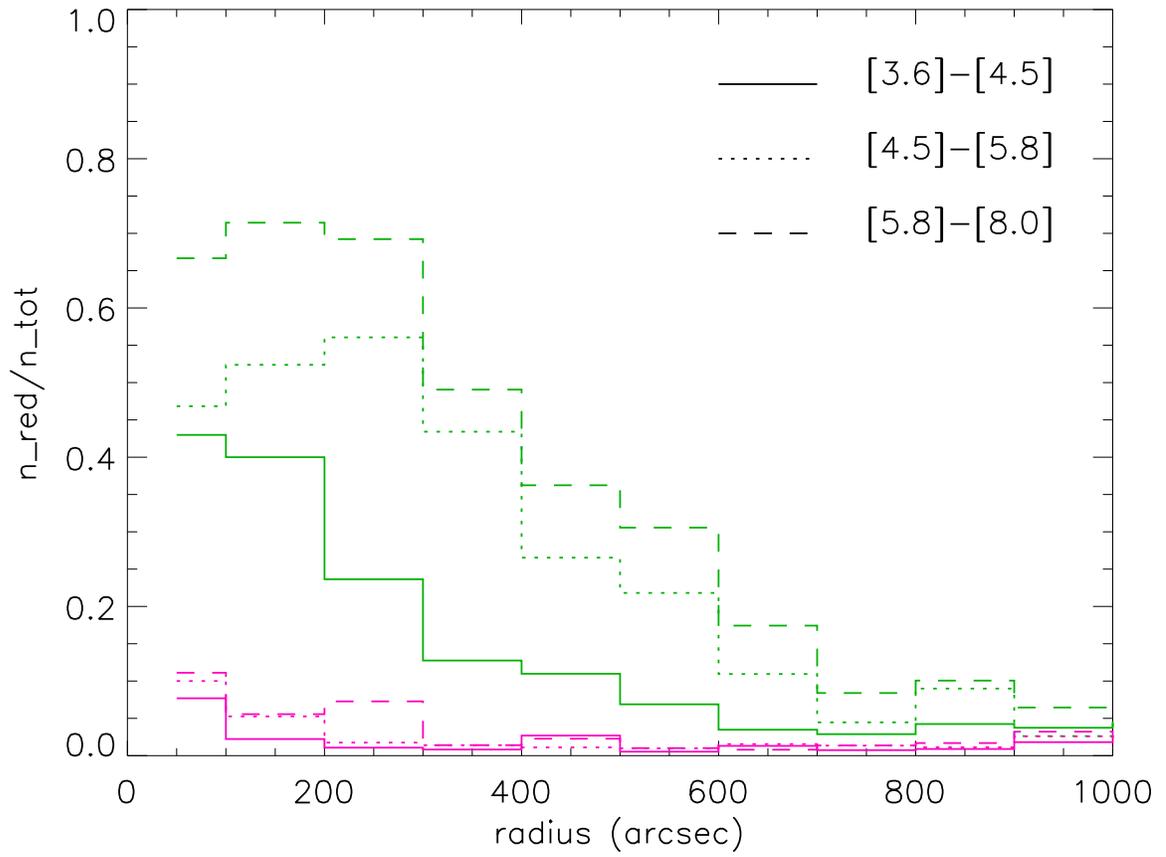} 
\caption{\label{radial} The green lines show the fraction of red stars as a function
of radius in RCW~49 for three IRAC colors.  The
pink lines show the fraction of red stars as a function of radius for the field 0.75\arcdeg\ north of RCW~49.  
}
\end{figure}

\begin{figure}
\plotone{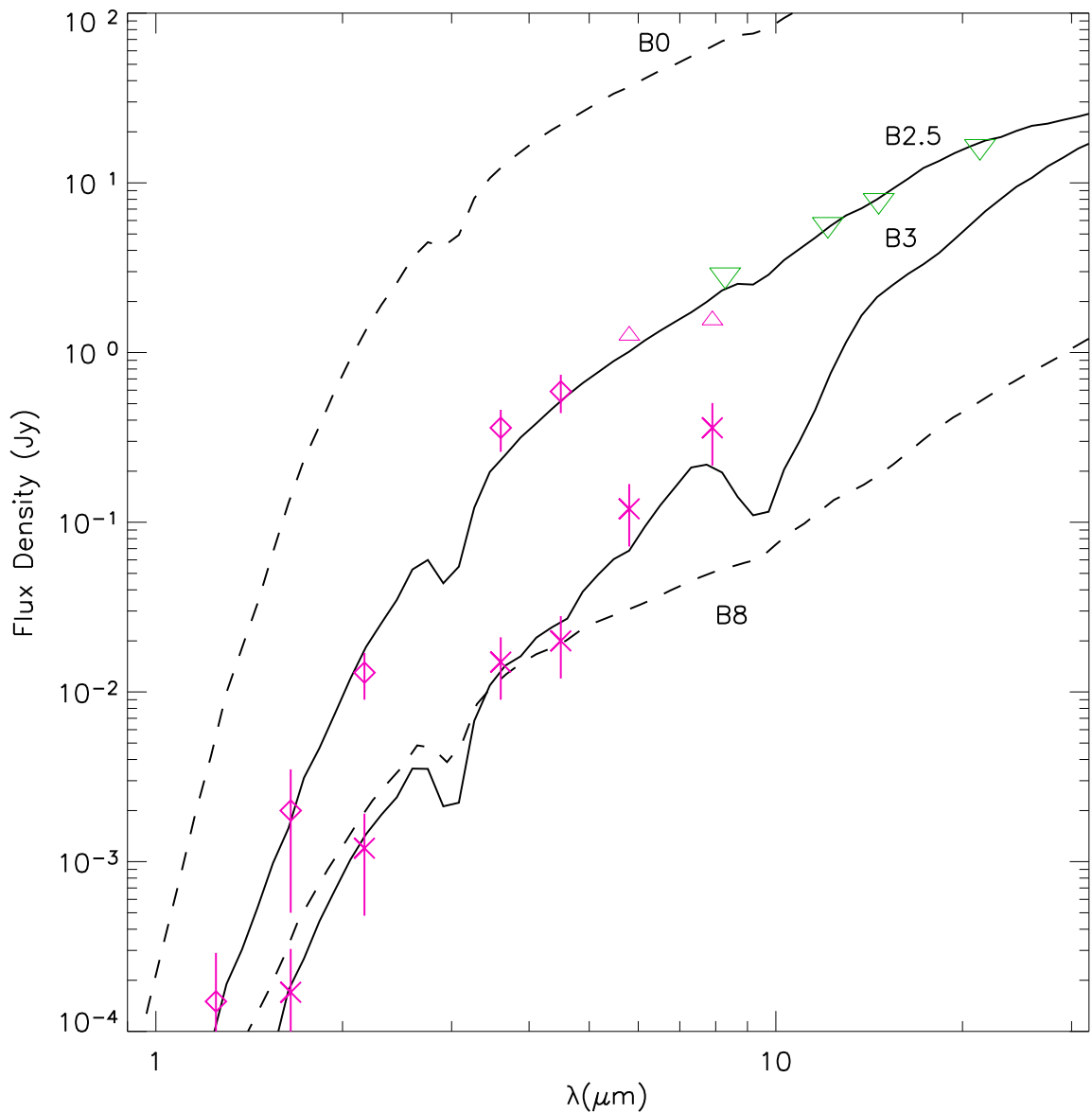}
\caption{\label{sed} Spectral energy distributions for the two
marked YSOs in the Region 2 (see Figure~\ref{regions}).
The bright YSO (left arrow in Figure~\ref{regions}) is
indicated by pink diamonds with error bars and upward-pointing
triangles in the saturated bands.
The triangular-shaped YSO (right arrow in Figure~\ref{regions}) is shown as pink x's with error bars.
The  {\it MSX} points are
shown as green downward-pointing triangles.
Models are shown for embedded YSOs with various central source spectral
types, all
at inclination 70\arcdeg. 
}
\end{figure}

The large number of infrared excess sources in RCW~49 indicates that
star formation is ongoing in this giant \ion{H}{2} region.  We find that at least 300 stars brighter
than 13$^{th}$ magnitude in the [3.6] band are pre-main
sequence or young stars of mass $\gtrsim 2.5$~M$_{\sun}$.
Based on integration of a standard IMF, we
conclude that the YSOs in the RCW~49 complex comprise a stellar mass of  $\sim$4500~M$_{\sun}$.  
If the YSOs represent a second generation of
star formation, this suggests that the mass of the second generation
is about 15\% of the first.  The presence of YSOs of B
spectral type does indeed suggest that some star formation is very
recent.  Whether it is continuous or triggered by the W2 cluster is
uncertain.  The apparent excess of star formation near the radio/IR
ring (Figure~\ref{3col}) and several of the sites of star formation
near the ring and blowout region perimeters (Regions 2,3, and 6 in
Figure~\ref{regions}) suggest that energy input from W2 may have
triggered recent star formation.  On the other hand, Regions~4 and 5
appear to be forming stars in a relatively quiescent region of the
cloud.

The detailed nature of the star formation in RCW~49---whether triggered or
continuous, clustered or distributed, and the spatial distribution of
mass---requires further study by the 
wider astronomical community.  
The images of this field have been released with the opening of the public
archive; the GLIMPSE catalog and mosaic images
will be available to the community from the
{\it Spitzer} Science Center in the fall of 2004.

\acknowledgments We are grateful to Stephan Jansen for his invaluable work maintaining
the GLIMPSE computing network.  
Support for this work, part of the Spitzer Space Telescope Legacy
Science Program, was provided by NASA through Contract Numbers (institutions)
1224653 (UW), 1225025 (BU), 1224681 (UMd), 1224988 (SSI), 1242593 (UCB), 1253153 (UMn), 
11253604 (UWy), 1256801 (UWW)  by the Jet Propulsion Laboratory, California
Institute of Technology under NASA contract 1407.

\clearpage

\clearpage

\clearpage

\end{document}